\documentclass[aip,onecolumn]{revtex4}
\usepackage{graphicx}
\usepackage{epsfig}
\usepackage{bm}
\usepackage{amsmath}
\usepackage{amssymb}
\usepackage{wasysym}
\usepackage{epstopdf}
\usepackage{color}
\usepackage{xcolor}
\usepackage{epstopdf}
\usepackage{upgreek}
\usepackage{subfloat}
\usepackage{graphicx}
\usepackage{dcolumn}
\usepackage{bm}
\usepackage{latexsym}
\usepackage{subfloat}
\usepackage{subfig}
\usepackage{psfrag}
\usepackage{xcolor}
\usepackage{natbib}
\usepackage{epstopdf}
\usepackage{amsmath}
\usepackage{comment}
\graphicspath{{Figures//}}

\usepackage[colorlinks=false, bookmarks=true]{hyperref}

\begin{document}


\title[Wall forces on a sphere in a rotating liquid-filled cylinder]{Wall forces on a sphere in a rotating liquid-filled cylinder} 
\author{Yoshiyuki Tagawa}
\email{y.tagawa@utwente.nl}
\author{Jarich van der Molen}
\author{Leen van Wijngaarden}
\author{Chao Sun}
\email{c.sun@utwente.nl}

\affiliation{ 
Physics of Fluids Group, Faculty of Science and Technology, J. M. Burgers Centre for Fluid Dynamics, University of Twente, PO Box 217, 7500 AE Enschede, The Netherlands\\}
  
\date{\today}

\begin{abstract}
We experimentally study the behavior of a particle slightly denser than the surrounding liquid in solid body rotating flow. 
Earlier work revealed that a heavy particle has an unstable equilibrium point in unbounded rotation flows. 
In the confinement of the rotational flow by a cylindrical wall a heavy sphere with density 1.05 g/cm$^3$ describes an orbital motion in our experiments.
This is due to the effect of the wall near the sphere, i.e. a repulsive force ($F_W$).
We model $F_W$ on the sphere as a function of the distance from the wall ($L$): $F_W \propto L^{-4}$ as proposed by \citet{Takemura2003}. 
Remarkably, the path from the model including $F_W$  reproduce the experimentally measured trajectory. 
In addition during an orbital motion the particle does not spin around its axis, and we provide a possible explanation for this phenomenon.
\end{abstract}

\maketitle

\section{Introduction}
\label{sec:Introduction}
Forces acting on a sphere in various flows are of importance from a fundamental point of view and in applications (see the review article by \citet[]{Magnaudet2000}).
A natural approach for analyzing the forces in complex flows is to decouple the flow effect into those of typical flows, like a solid-body rotating flow.
In recent years the forces on particles lighter than the surrounding fluid were studied in a solid-body rotation flow around a horizontal axis (i.e. gravity is perpendicular to the rotation axis)  by \citet{Nierop2007, Bluemink2005, Bluemink2008, Bluemink2010, Rastello2009, Rastello2011}. 
The advantage of this system is that buoyant particles reach an equilibrium point, from which drag and lift coefficient are well determined over a wide range of dimensionless parameters.
Provided the equilibrium point is sufficiently far from the wall(s) bounding the rotating flow, this can be considered as unbounded.
In this way drag and lift forces as well as particle spin have been measured at moderate Reynolds numbers  and reported in the above mentioned papers.
A particle heavier than surrounding liquid spirals outward in rotating flow, i.e., it has an unstable equilibrium point.
Such orbits were calculated by \citet{Roberts1991} for such small particles that Stokes flow can be assumed.
They also took no lift forces into account.


The effect of the wall is important but less intensively studied, especially at moderate-to-high Reynolds numbers.
It has been investigated mainly in the situation of a sphere moving parallel to a flat plate or rotating near a plane boundary.
The influence of a vertical flat wall on a spherical rising bubble is studied experimentally by \citet{Takemura2003} and numerically by \citet{Zeng2005} for Reynolds number less than 100.
\citet{Takemura2003} discussed two hydrodynamical mechanisms of wall interaction, one due to the vorticity generated at the bubble surface and the other to the irrotational dipole associated with the bubble motion. They proposed empirical correlations based on the strength of these two mechanisms, leading to practical expressions of the lift force as a function of Reynolds number and the distance between the wall and the particle. 
\citet{Liu2010} studied the effect of the spin of the sphere in the presence of no-slip planar boundaries.
They found that for small sphere-wall gap widths and Reynolds numbers $\sim$ 1 viscous effect plays a role causing repulsive forces while larger Reynolds numbers $\sim$ 100 cause a Bernoulli effect in the gap, which may turn the forces into attractive ones.
 
When it comes to rotating flow, the wall effect is studied for very viscous flow.
\citet{Ashmore2005} investigated the interaction of the wall with a dense steel particle in the Stokes regime. 
\citet{Mullin2005} found that with a very heavy sphere in a highly viscous rotating fluid three regimes can be distinguished as a function of the cylinder rotating speed, which we will discuss in \S~\ref{sec:Results}.

In this paper we experimentally investigate the trajectories of particles, slightly heavier than the rotating fluid in which the particles are introduced.
In particular we are interested in the interaction with the cylinder wall for higher Reynolds number O(10$^3$).
The translational and rotational motion of the spheres are observed with help of Particle Tracking Velocimetry  (PTV) and the orientation detection software~\cite{Zimmermann2011}, respectively. 
We use Particle Image Velocimetry (PIV) to record the velocity distribution in the undisturbed as well as in the disturbed fluid 

The outline of this paper is as follows. In \S~\ref{sec:forces} the equation of motion of a particle is introduced. The experimental setup is described in \S~\ref{sec:Experimental}. The results  are shown in \S~\ref{sec:Results}, followed by conclusive remarks in \S~\ref{sec:concl}.


\section{Forces acting on a particle}
\label{sec:forces}
 
\begin{figure}
	\centering
\includegraphics[width=0.4\textwidth]{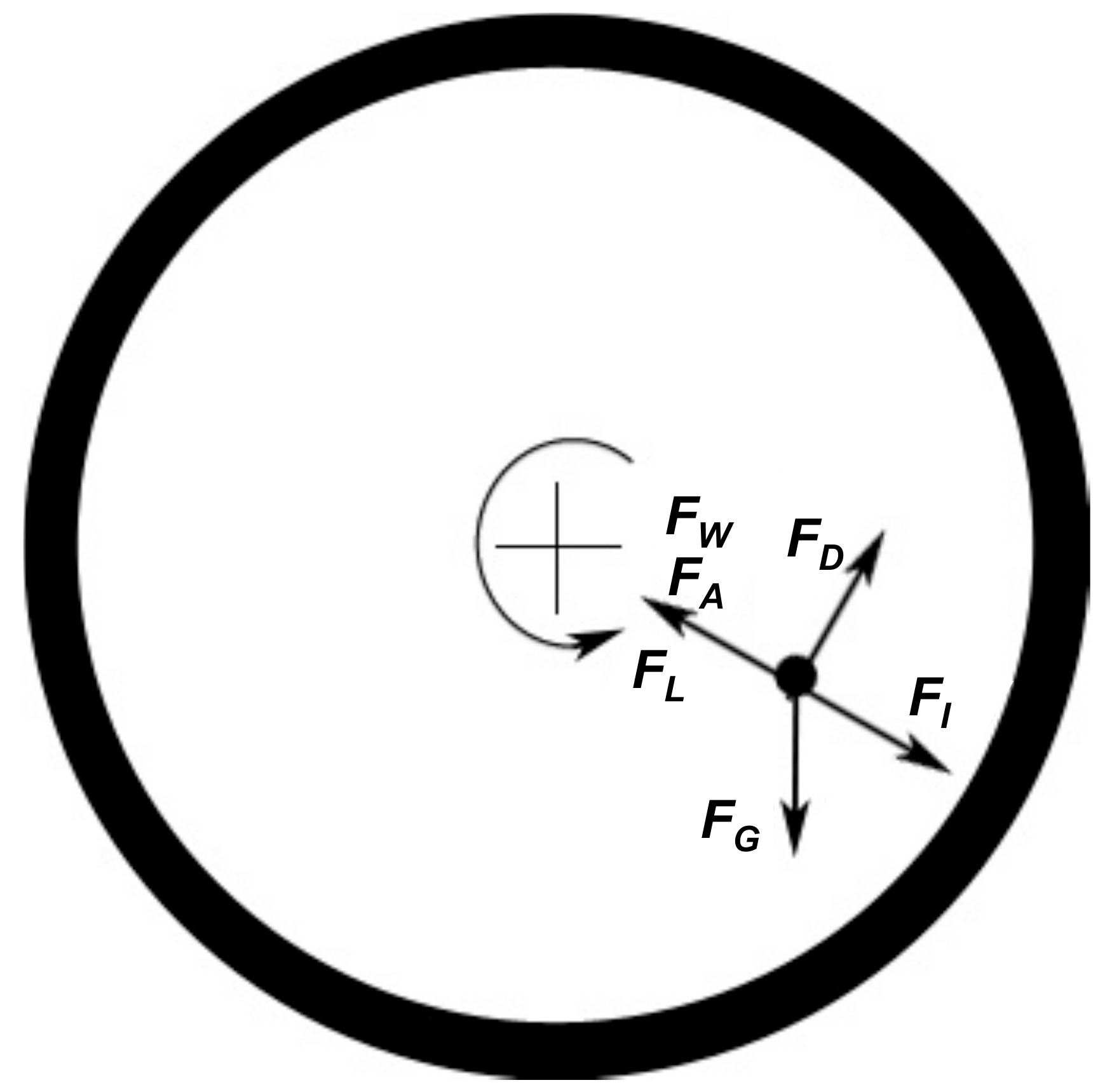}
	\caption{Forces acting on a particle: $\boldsymbol{F_A}$ the added mass force, $\boldsymbol{F_D}$ the drag force, $\boldsymbol{F_L}$ the lift force, $\boldsymbol{F_G}$ the gravity force or body force, $\boldsymbol{F_I}$ the force due to acceleration of the flow, and $\boldsymbol{F_W}$ the wall repulsive force. }
	\label{fig:forces}
\end{figure}

The force balance on a sphere in the present condition is 
\begin{equation}
\rho_pV_p \frac{d\boldsymbol{u}}{dt} =\boldsymbol{F_A}+\boldsymbol{F_D}+\boldsymbol{F_L}+\boldsymbol{F_G}+\boldsymbol{F_I}+\boldsymbol{F_W},
\end{equation}
where $\rho_p$ is the density of the particle, $V_p$ the volume of the particle, $\boldsymbol{u}$ the particle velocity, $\boldsymbol{F_A}$ the added mass force, $\boldsymbol{F_D}$ the drag force, $\boldsymbol{F_L}$ the lift force, $\boldsymbol{F_G}$ the gravity force or body force, $\boldsymbol{F_I}$ the force due to acceleration of the flow, and $\boldsymbol{F_W}$ the wall repulsive force. All forces in the rotating flow are shown in Fig.~\ref{fig:forces}.
The operators $d/dt$ and $D/Dt$ are material derivatives going with the particle and with the fluid, respectively.
Owing to the fact that the present configuration is quite similar to that of light particles investigated by \citet{Bluemink2010}, we can adopt expressions for $\boldsymbol{F_A}$, $\boldsymbol{F_D}$, $\boldsymbol{F_L}$, $\boldsymbol{F_G}$, and $\boldsymbol{F_I}$ from \citet{Magnaudet2000, Mazzitelli2003}. 
The full differential equation with all relevant variables is
\begin{multline}
\rho_p V_p \frac{d\boldsymbol{u}}{dt} = \rho V_pC_A \left(\frac{D\boldsymbol{U}}{Dt}-\frac{d\boldsymbol{u}}{dt} \right)  + \rho V_p C_L(\boldsymbol{U}-\boldsymbol{u})\times(\boldsymbol{\nabla} \times \boldsymbol{U})+\frac{1}{2}\rho A_pC_D|\boldsymbol{U}-\boldsymbol{u}|(\boldsymbol{U}-\boldsymbol{u}) + (\rho_p - \rho)V_p\boldsymbol{g} + \rho V_p\frac{D\boldsymbol{U}}{Dt} +\boldsymbol{F_W},
\label{eq:EoM}
\end{multline}
where $\rho$ is density of the fluid, $C_A$ added mass coefficient ($C_A$=1/2 for the sphere case), $\boldsymbol{U}$ the velocity of the fluid, $C_L$  the lift coefficient normalized with the vorticity $\boldsymbol{\nabla} \times \boldsymbol{U}$, $A_p$  the cross sectional area of the sphere, and $\boldsymbol{g}$  the gravity acceleration. 


We have not found results on the magnitude of a wall induced force $\boldsymbol{F_W}$ in case of a curved wall, as is ours.
Since the radius of the sphere is small with respect to the drum radius, we may confidently use results for a flat wall.
Recent work, relevant for our case, is given by \citet{Takemura2003} and \citet{Zeng2005}.
From these work it is clear that there are two mechanisms which contribute to $\boldsymbol{F_W}$.
The first is the vorticity distribution in the wake behind the sphere.
This diffuses outward, but this process is asymmetric due to the presence of the wall.
It leads to a wall force away from the wall.
On the other hand the accelerated flow through the gap between the sphere and the wall produces an attractive force.
The first mechanism is dominant over a wide range of Reynolds numbers.
\citet{Takemura2003} did experiments at Reynolds number of order 10$^2$ and over a large range of the ratio between distance from a wall and sphere radius, $L/R$.
They summarized their results as follows.

\begin{subequations}
\begin{equation}
F_W = C_WA_p\rho u^2/2,
\end{equation}
where
\begin{equation}
C_W = C_{W0}(L^*)a^2(Re)(L/\gamma R)^{g(Re)}
\end{equation}
with
\begin{equation}
\gamma \approx 3.0,\qquad g(Re)\approx -2.0\tanh{(0.01Re)},\qquad a(Re)\approx 1 + 0.6Re^{1/2}-0.55Re^{0.08} 
\end{equation}
\begin{equation}
C_{W0}(L^*) = \left\{
\begin{array}{ll}
(9/8+5.78\cdot 10^{-6})L^{* 4.58}\beta^2\exp{(-0.292L^*)} & \mbox{for   }  0<L^*<10 \\
8.94\beta^2L^{* -2.09} & \mbox{for }10 \le L^* < 300,
\end{array}
\right.
\end{equation}
\label{eq:F_W}
\end{subequations}
where $\beta$ = 1 for a rigid sphere. The dimensionless separation $L^*$ = $LU/\nu$, the Reynolds number $Re = Ud_p/\nu$ based on the particle diameter $d_p$ and the oncoming fluid velocity at the particle location $U$.

\citet{Zeng2005} performed numerical calculations for the situation of the experiments by \citet{Takemura2003}, and found good agreement with Eq.~(\ref{eq:F_W}a-d) in the relevant parameter range.
Although our Reynolds numbers are larger than those in the experiments of \citet{Takemura2003}, we assume that the same trends with Reynolds number.

We choose cylindrical coordinates ($r, \phi, z$) with $z$ along the rotation axis.
Since during the experiments, to be described in the next section, the particle always stays in a plane perpendicular to the rotation axis, the momentary position is sufficiently described by $\boldsymbol{r} = r\boldsymbol{e}_{r} + \phi\boldsymbol{e}_{\phi}$, where $\boldsymbol{e}_{r}$ and $\boldsymbol{e}_{\phi}$ are unit vectors in radial and azimuthal directions respectively.
With a constant angular velocity $\omega$, the flow velocity is
\begin{equation}
\boldsymbol{U(r)} = \omega r \boldsymbol{e_{\phi}}.
\end{equation}
In terms of this coordinate system, Eq.~(\ref{eq:EoM}) becomes
\begin{subequations}
\begin{align}
\ddot{r}&=\left(\frac{1}{\rho_p+C_A \rho}\right)\left(\left(\rho_p+C_A\rho\right)r\dot{\phi}^2+\left(\rho-\rho_p\right)g\sin{\phi}  -\frac{3}{8}R_pC_D\rho\sqrt{\dot{r}^2+\left(r\left(\dot{\phi}-\omega\right)\right)^2}\dot{r} +2C_L\rho \omega r \left(\omega-\dot{\phi}\right)\right. \nonumber \\ & \qquad \left.{} -\rho r\omega^2\left(C_A+1\right) +F_{W}\right)
\label{eq:rddot}
\end{align}
for the radial direction. 
For the azimuthal direction, we obtain
\begin{align}
\ddot{\phi}&=\left(\frac{1}{r\left(\rho_p+C_A\rho\right)}\right)\left(-2\left(\rho_p+C_A\rho\right)\dot{r}\dot{\phi}+\left(\rho-\rho_p\right)g\cos{\phi}+\frac{3}{8R_p}C_D\rho\sqrt{\dot{r}^2+\left(r\left(\dot{\phi}-\omega\right)\right)^2}r\left(\omega-\dot{\phi}\right)+2C_L\rho\omega\dot{r}\right).
\label{eq:phiddot}
\end{align}
\label{eq:rphiddot}
\end{subequations}
The dots denote differentiation with respect to time.

\section{Experimental set-up}
\label{sec:Experimental}

\begin{figure}
	\centering
\includegraphics[width=.7\textwidth]{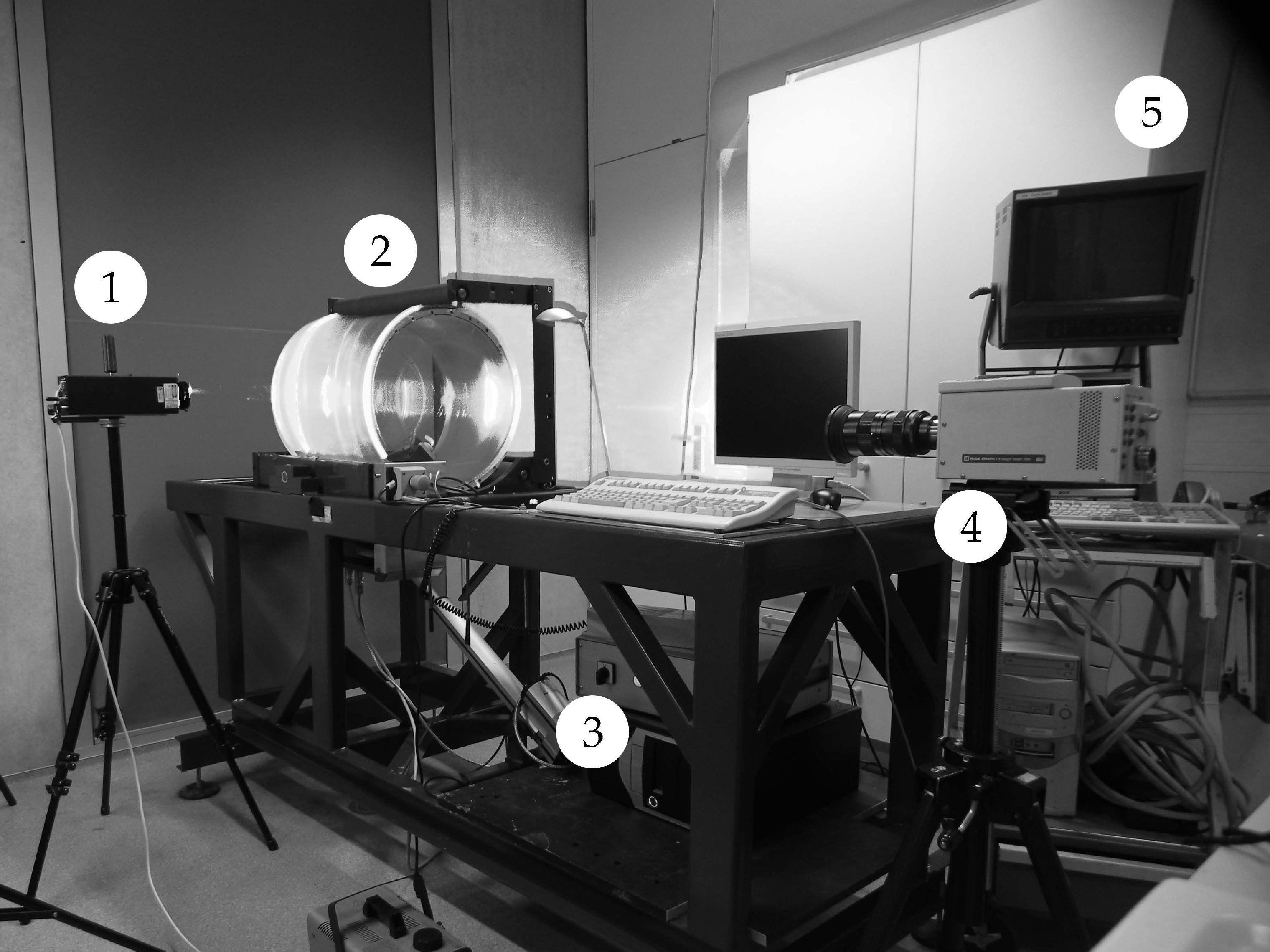}
	\caption{Experimental setup. 1. Laser, 2. Drum, 3. Drum controller, 4. Camera, 5. Camera controller}
	\label{fig:setup}
\end{figure}

Figure~\ref{fig:setup} shows the experimental setup. 
The cylindrical acrylic drum is 500 mm long and has a radius of 250 mm with 15 mm thick plastic walls and lids. 
Its axisymmetric axis is horizontal. 
Two steel rods with rubber coatings support it and one of them rotates driven by an AC servo motor. 
We vary the frequency of the drum $f_d$ in the range of 0-2 Hz. 
The drum is filled with de-ionized water with density 1.00 g/cm$^3$. 
We use slightly heavier polysterene spheres with density 1.05 g/cm$^3$. 
Their size ranges from 10 to 60 $\pm$ 0.01 mm. 
The adsorption of de-ionized water by polysterene is less than one procent by volume. 

Particle trajectories are recorded by using a high-speed camera (Kodak 2000, Redlake Co., USA /  Photron SA.1, Photron Co., Japan) with a lens (Fujinon TV Zoom lens).
Recording speeds are between 60 frames per second (fps) and 500 fps. 
In general, the measurement time is about 6 minutes which corresponds to more than 20 particle trajectory cycles.
To ensure stable orbits, the drum is left rotating together with the inserted particle for more than 45 minutes before starting each measurement with set drum frequency. 

\begin{figure}
	\centering
	\includegraphics[width=.95\textwidth]{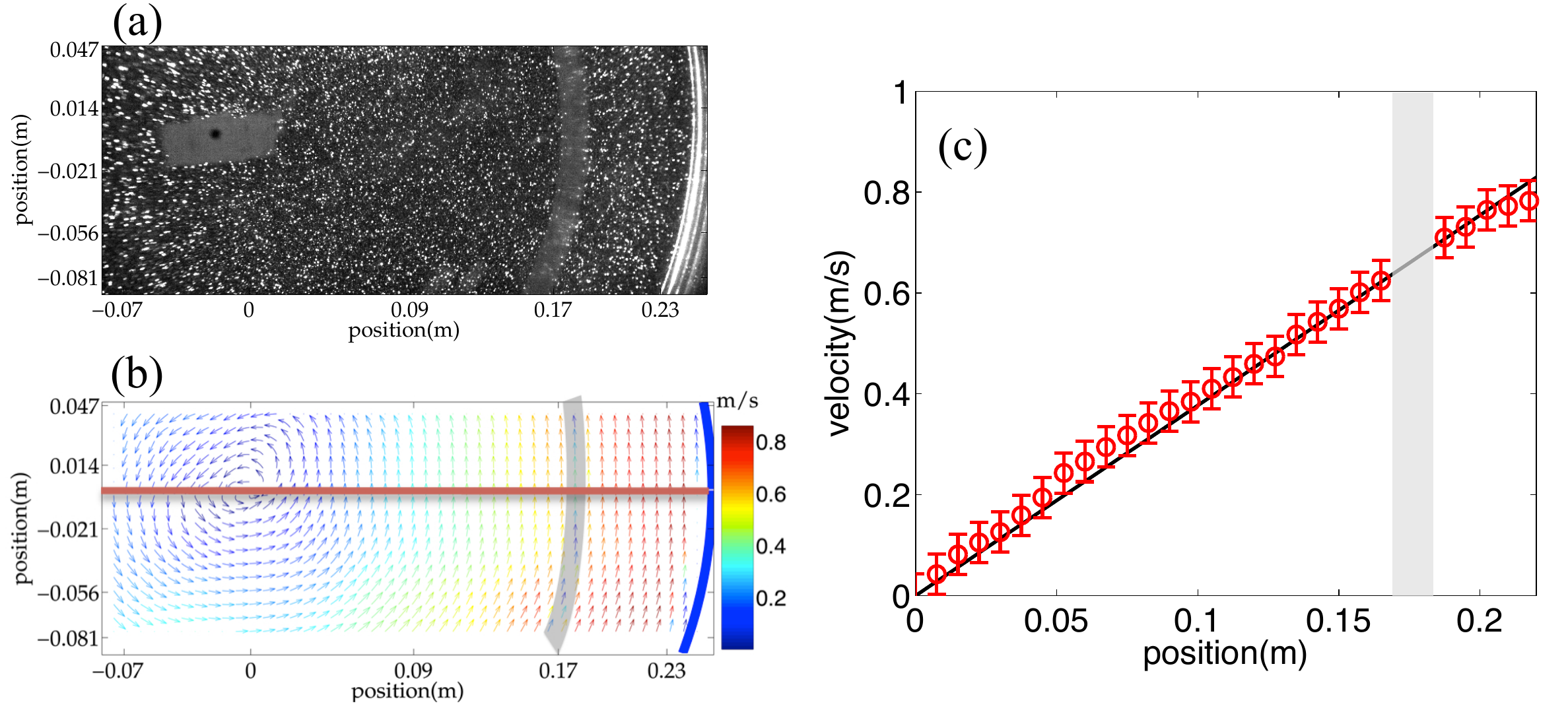}	
	\caption{Solid-Body rotational flow at a drum frequency of 0.60Hz: (a) A snapshot of the PIV tracer particles. (b) Flow field. The scratched region is highlighted in gray color. (c) Velocity vs. position on a horizontal line (red line in the left figure).  }
	\label{fig:SolidBodyRotationalFlow}
\end{figure}

The recorded images are analyzed with PTV software. 
The particle center detected by the use of a circular Hough transform method represents its position. 
The position of the wall of the drum is determined using particle reflection images. 
As shown in Fig.~\ref{fig:SolidBodyRotationalFlow}(a), the lid of the drum has scratched areas at the radial position of from 15.5 cm to 17 cm.
When the particle locates In the scratched area, we determine the particle position using linear interpolation. 

The flow in the drum is measured using PIV technique with a laser sheet (Lasiris Magnum 2, StockerYale on Coherent inc., Canada). 
The laser-beam illuminates a cross-sectional area of the drum. 
The flow filed at the drum frequency $f_d$=0.60 Hz, without a sphere, is recorded at 1000 fps and an exposure time of 1 $\mu$s. 
Figure~\ref{fig:SolidBodyRotationalFlow}(b) shows the flow filed by averaging 250 frames. 
This shows a solid-body rotation. 
The velocity along the horizontal line (solid red line in Fig.~\ref{fig:SolidBodyRotationalFlow}b) is shown in Fig.~\ref{fig:SolidBodyRotationalFlow}(c) with the theoretical line of the solid body rotation. 
The scratched area of 15.5 to 17 cm in radial position is indicated in gray color, where the mean velocity is not measurable.
The experimental result agrees well with the predicted line within error bars, confirming that the flow is solid body rotation.

\section{Results and discussions}
\label{sec:Results}


\begin{figure}
	\centering
\includegraphics[width=0.35\textwidth]{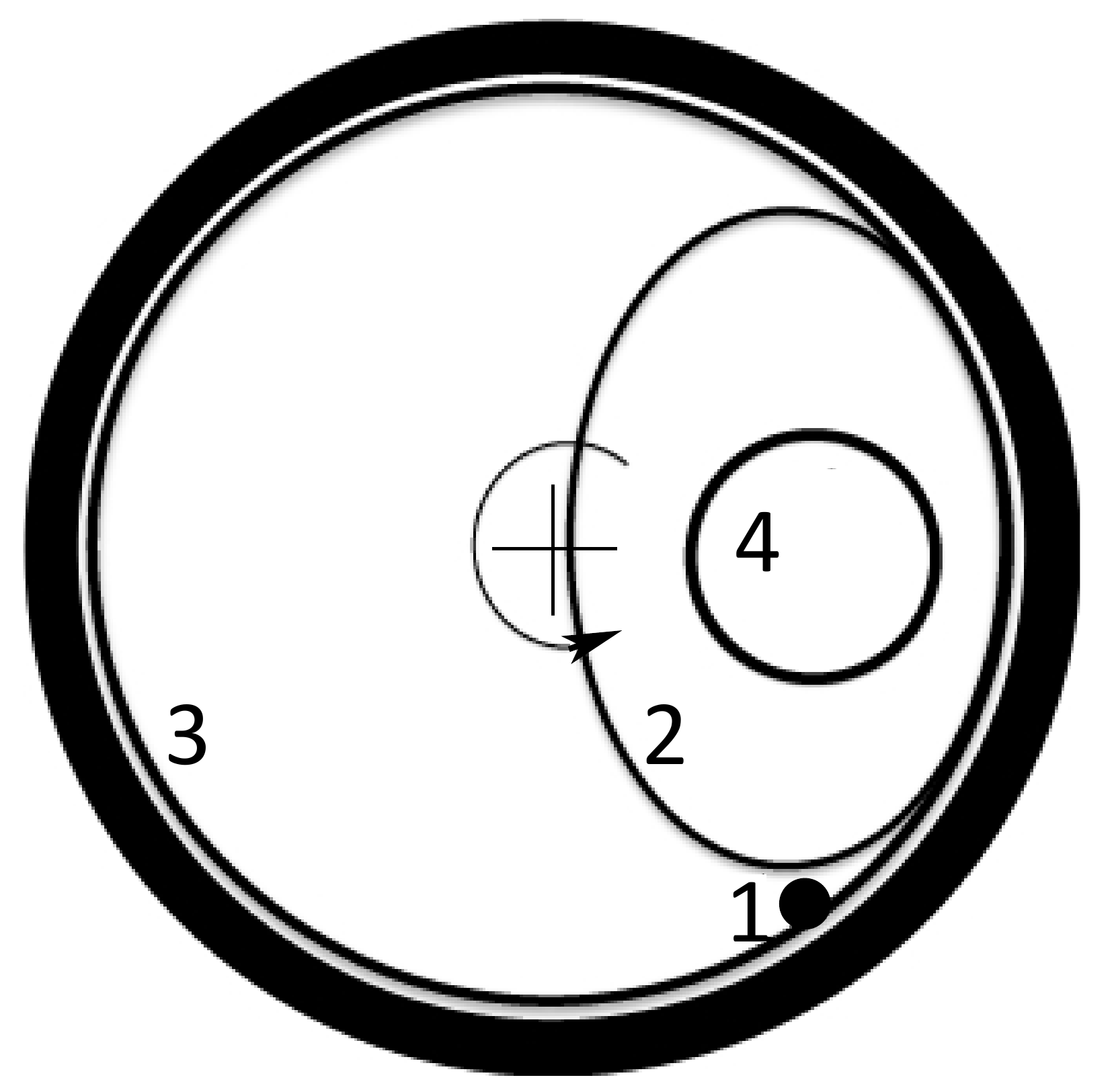}
	\caption{Four regimes of a heavy particle trajectory: (1) Fixed-point regime, (2) cascading regime, (3) fixed solid body rotation regime, and (4) suspension regime.}
	\label{fig:regimes}
\end{figure}

We describe what a particle, radius 7 mm, density 1.05 g/cm$^3$ does when immersed in the rotating flow and the drum frequency $f_d$ is increased from zero to 1.8 Hz.
Much of its behavior is similar to that of heavy particles as described by \citet{Mullin2005}, mentioned in the introduction. 
For $f_d<$0.07 Hz the particle rolls along the drum wall at a fixed position as regarded in the laboratory frame.
This is what \citet{Mullin2005} call the fixed point regime shown as trajectory (1) in Fig.~\ref{fig:regimes}.
The other regimes mentioned by \citet{Mullin2005}  are found as well.
For 0.12 Hz$<f_d<$1.2 Hz the particle touches the drum wall in a part of the orbit and falls down in the remainder.
This is the cascading regime (trajectory (2) in Fig.~\ref{fig:regimes}).
For $f_d>$1.2 Hz the particle sticks on the wall of the drum and rotates with it.
This is the fixed solid body regime (trajectory (3) in Fig.~\ref{fig:regimes}). 

%

However, we found that our particle, \textit{slightly} heavier than the fluid, floats through the drum, without contacting the drum wall, at drum frequencies between 0.07 and 0.11 Hz.
We call this the suspension regime shown as trajectory (4) in Fig.~\ref{fig:regimes}.
Clearly this is possible by the presence of the wall.
In the remaining part of this section we focus on the particle motion in this regime.

\begin{figure}
	\centering
	\includegraphics[width=0.6\textwidth]{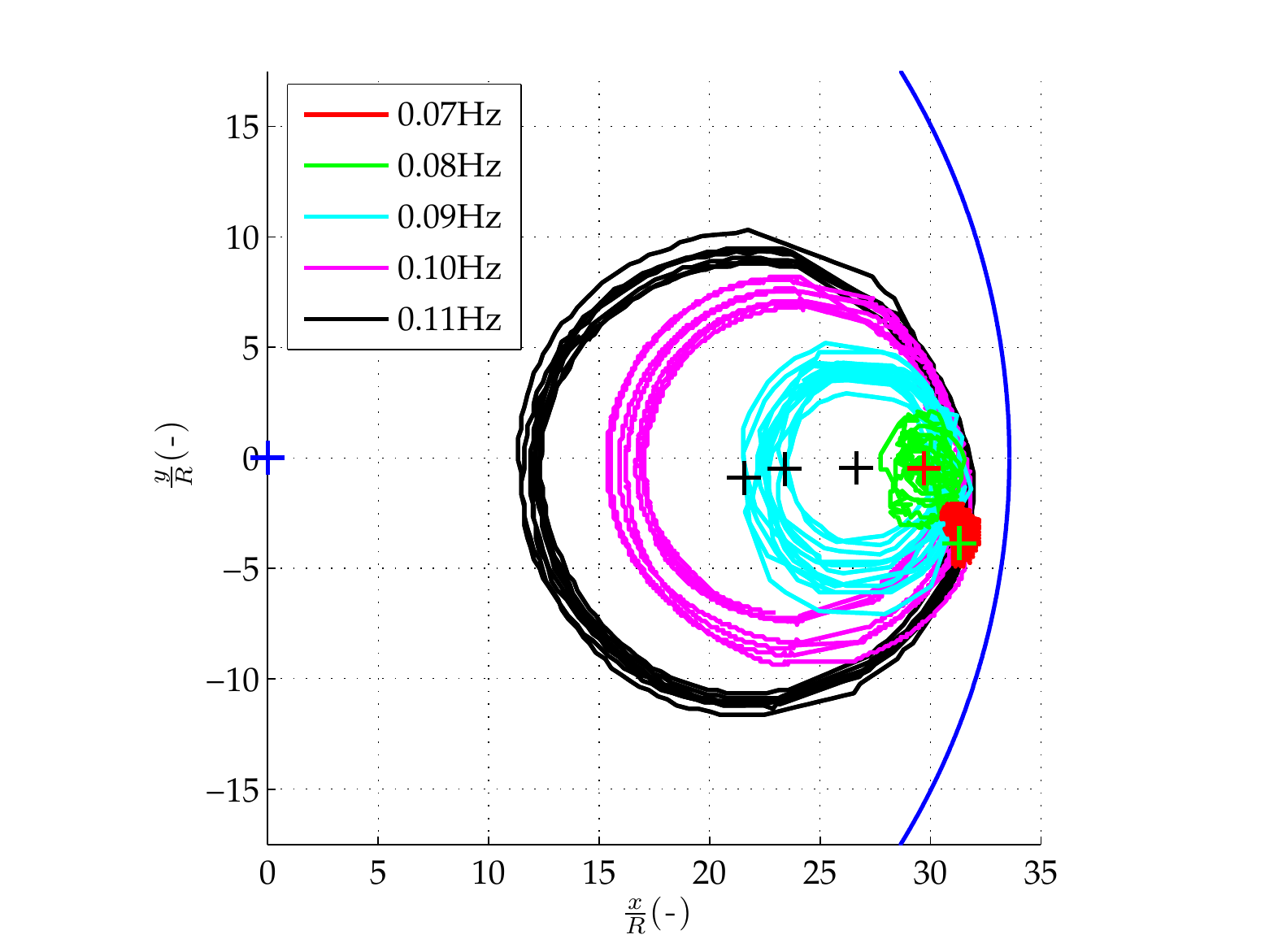}
	\caption{Particle position of a 7mm polystyrene particle in a drum for frequency 0.07 - 0.11 Hz.}
	\label{fig:ParticleMeasuredPosition}
\end{figure}

\begin{figure}
	\centering
	\includegraphics[width=0.5\textwidth]{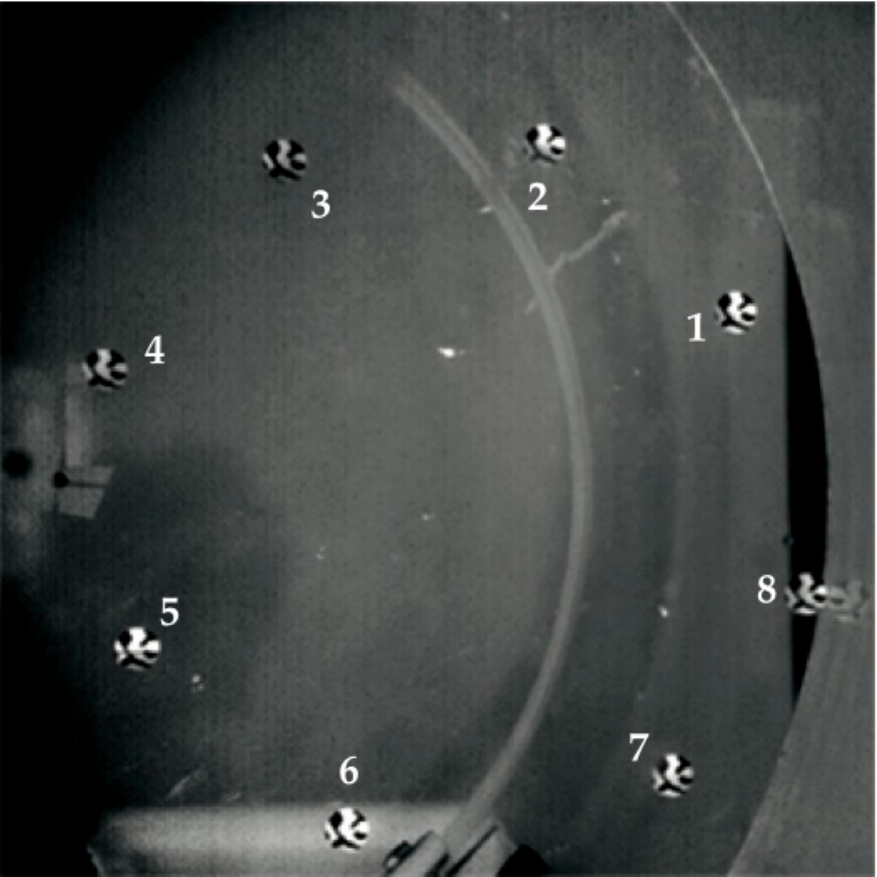}
	\caption{Images of a slightly heavy particle with black and white texture in a drum}
	\label{fig:NonRotatingParticles}
\end{figure}

The positions of the particle are plotted in Fig.~\ref{fig:ParticleMeasuredPosition} for several drum frequencies $f_d$.
The coordinates of the particle positions in Fig.~\ref{fig:ParticleMeasuredPosition} are made dimensionless with help of the particle radius. 
The trajectories show almost perfect circles.
The minimum particles distance to the wall is $\sim$ 1 for all cases. 
Their radius increases with increasing $f_d$ and their central positions ($r_e$, $\phi_e$), shown as cross symbols in Fig.~\ref{fig:ParticleMeasuredPosition}, depart from the wall with $f_d$.
The coordinates $r_e$ and $\phi_e$ mark the equilibrium position.
This position apparently is not stable in the sense that upon a displacement the particle returns to the original position.
However, it does not spiral outward or inward as in the study of \citet{Roberts1991}, but remains in orbit about ($r_e$, $\phi_e$).
The explanation for this lies in the fact that in the analysis of \citet{Roberts1991} there is no lift force.
At the equilibrium position ($r_e$, $\phi_e$) in our case $C_D$ and $C_L$ can be calculated by taking $\ddot{r}=\dot{r}=\ddot{\phi}=\dot{\phi}=0$ in eq.~\ref{eq:rphiddot} as:
\begin{subequations}
\begin{equation}
C_D = \frac{4}{3}\frac{gd\cos{\phi_e}}{r_e^2\omega^2}
\end{equation}
\begin{equation}
C_L = \frac{1}{2}(1+C_A-\frac{g\sin{\phi_e}}{r_e\omega^2}).
\end{equation}
\label{eq:equilibrium}
\end{subequations}
From Eq.~\ref{eq:rphiddot} it follows that in the $r=r_e$, $\phi = 0$ situation the drag is vertical and balances effective gravity.
Further the central force exerted by the pressure gradient in the rotating fluid is balanced by the lift force.
When now the particle undergoes a vertical displacement in r-direction, it meets with a drag which in excess of the buoyancy force and hence moves upward.
Then, it acquires a $\dot{\phi}$, by which the lift is reduced and it is pushed inward by the pressure force.
In the case of \citet{Roberts1991} there is no lift force and so the pressure gradient force is not balanced in $r=r_e$, $\phi = 0$.
A picture of an orbit as measured by us is shown in Fig.~\ref{fig:NonRotatingParticles}.
It is clear that between the points marked as 1, 2 etc. the angular velocity varies slightly.
It is also remarkable that, as the particle in the marked positions shows, the particle does not rotate about an axis through its own center.

We are, of course, interested whether the Eq.~(\ref{eq:rphiddot}) are able to reproduce the observed orbits and the absence of spinning motion.
The resulting range of the Reynolds number is 1400-1500.
In this $Re$ regime, the expression of Eq.~(\ref{eq:F_W}c-d) can be reduced to $g(Re)\approx -2.0$, $C_{W0}(L^*) \propto L^{-2}$.
Thus, the wall repulsive force may be written as
\begin{equation}
F_W \approx \chi L^{-4},
\label{eq:F_Wapprox}
\end{equation}
where $\chi$ is a constant. 
We used in Eq.~(\ref{eq:rddot}) the expression~(\ref{eq:F_Wapprox}) for the wall force $F_W$.
In addition we used $\chi$ as a fitting parameter.
We start with analyzing the trajectories.
\begin{figure}
	\centering
	\includegraphics[width=0.35\textwidth]{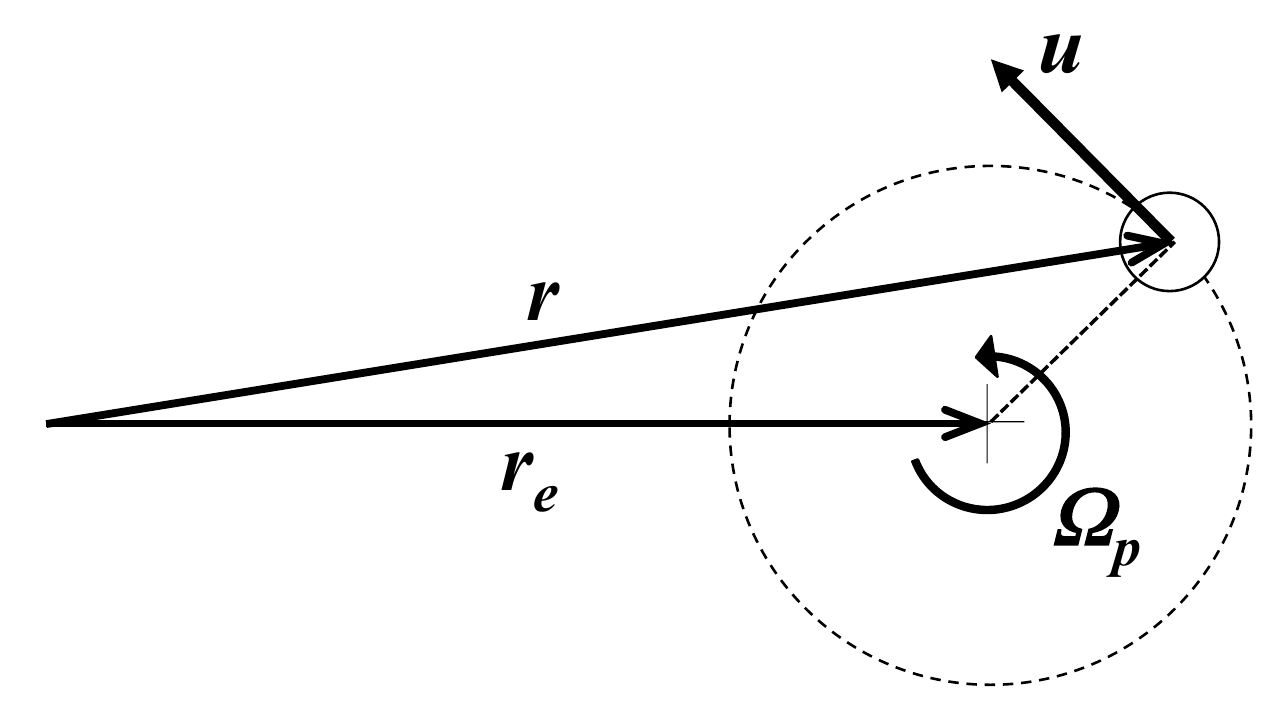}
	\caption{The equilibrium radius vector $\boldsymbol{r_e}$, the momentary position of the sphere $\boldsymbol{r}$, and the particle velocity $\boldsymbol{u}$. The particle turns around $\boldsymbol{r_e}$ with momentary angular velocity $\boldsymbol{\Omega_p}$. The dashed black line represents the particle circular orbit. The cross marker shows the equilibrium point.}
	\label{fig:Vectors}
\end{figure}
We denote the radius vector of the orbit center, measured from the center of the drum, with $\boldsymbol{r_e}$ and of the momentary position of the sphere with $\boldsymbol{r}$ as shown in Fig.~\ref{fig:Vectors} .
The particle turns around $\boldsymbol{r_e}$ with instantaneous angular velocity $\boldsymbol{\Omega_p}$ which varies along the orbit. 
So its velocity is

\begin{equation}
\boldsymbol{u} = \boldsymbol{\Omega_p}x(\boldsymbol{r}-\boldsymbol{r_e}).
\end{equation}

This is the velocity as measured in the laboratory frame.
The velocity relative to the surrounding fluid which rotates about the center of the drum with angular velocity $\boldsymbol{\omega}$, is

\begin{equation}
\boldsymbol{u}-\boldsymbol{\omega}x\boldsymbol{r} = (\boldsymbol{\Omega_p}-\boldsymbol{\omega})x(\boldsymbol{r}-\boldsymbol{r_e})-\boldsymbol{\omega}x\boldsymbol{r_e}
\label{eq:RelativeVel}
\end{equation}

\begin{figure}
	\includegraphics[width=0.6\textwidth]{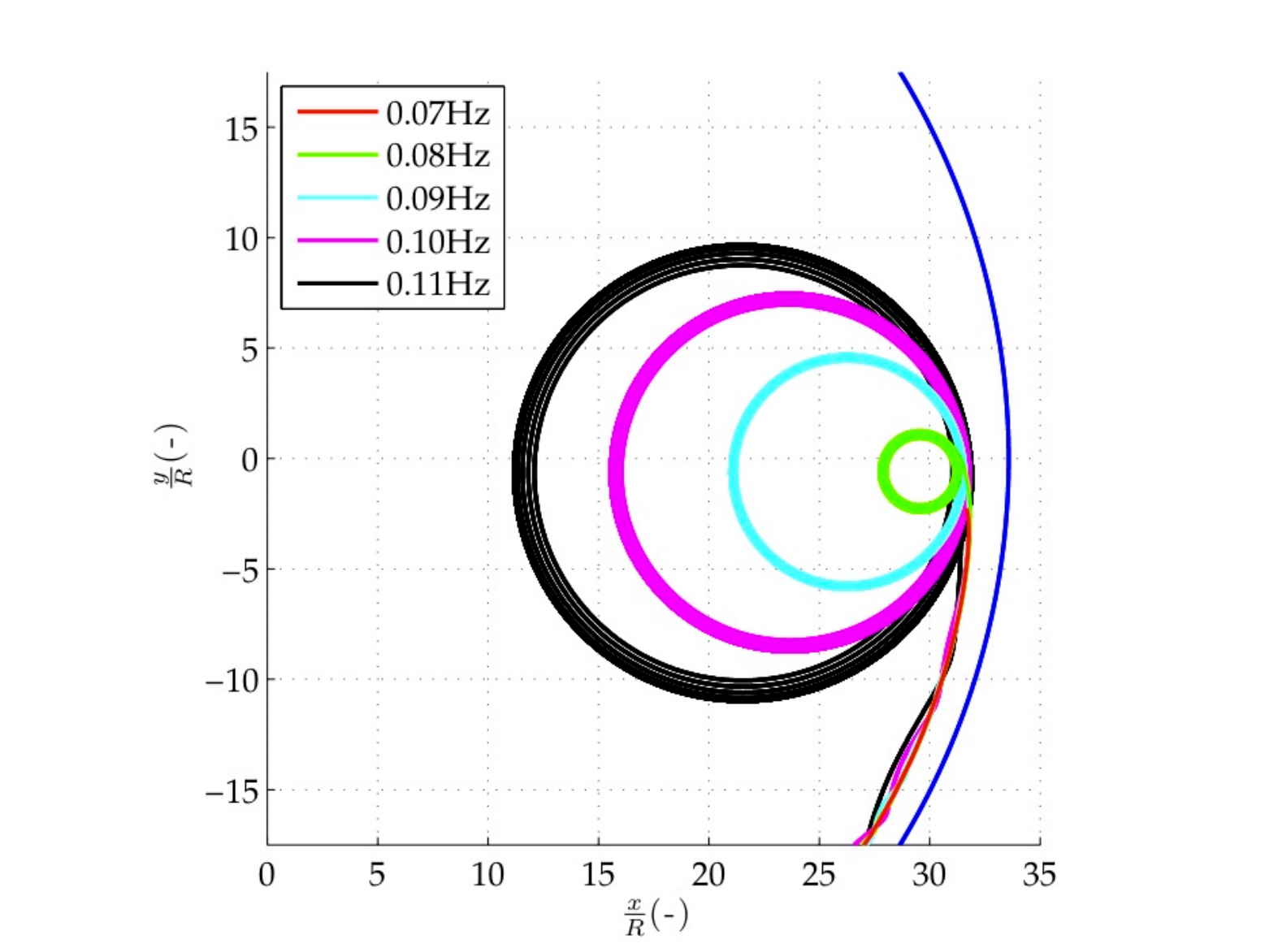}
	\caption{Paths reproduced by Eq.~(\ref{eq:rphiddot}) for the drum frequency 0.07 - 0.11 Hz.}
	\label{fig:ODE}
\end{figure}

\begin{figure}
	\includegraphics[width=0.5\textwidth]{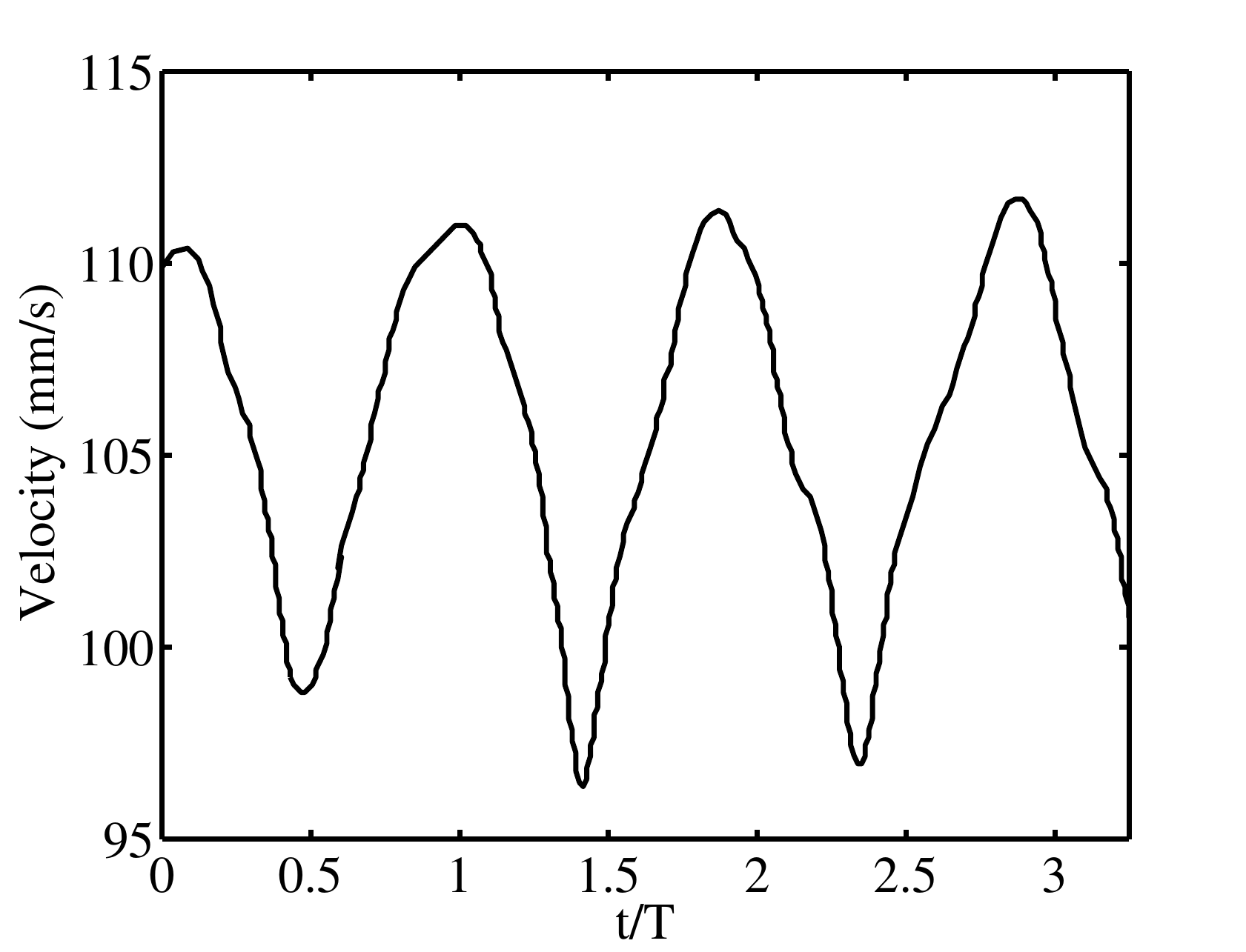}
	\caption{Time normalized by the drum time cycle vs. the magnitude of the particle velocity.}
	\label{fig:RelativeVelocity}
\end{figure}

The velocity $-\boldsymbol{\omega}x\boldsymbol{r_e}$ is constant along the orbit and it directed along the negative y-axis in Fig.~\ref{fig:ParticleMeasuredPosition}.
We can make an estimate for this from the measurements of the orbits and the velocities.
We take the case with the following data: $R = 7\cdot 10^{-3}$ m; $f_d = 0.11$ Hz.
The orbit is displayed in Fig.~\ref{fig:ODE} and the velocity relative to the rotating frame in Fig.~\ref{fig:RelativeVelocity}.

We write Eq.~(\ref{eq:RelativeVel}) as

\begin{equation}
\boldsymbol{u} = -\omega r_e\boldsymbol{e_y} + \Delta\Omega\sigma\boldsymbol{e_{\theta}} = < \boldsymbol{u}>+\boldsymbol{u'},
\label{eq:udecompose}
\end{equation}
where $\Delta\Omega = |\boldsymbol{\Omega_p}-\boldsymbol{\omega}|$ and $\sigma = |\boldsymbol{r}-\boldsymbol{r_e}|$.

From Fig.~\ref{fig:ODE} we obtain $r_e$ = 0.15 m and hence $<u>$ should be
\begin{equation}
<u> = 2\pi f_d r_e = 104\cdot10^{-3} m/s.
\end{equation}

In Fig.~\ref{fig:RelativeVelocity} the maximum and minimum velocity are $111\cdot10^{-3}$ m/s and $99\cdot10^{-3}$ m/s, respectively and hence the average is $104\cdot10^{-3}$ m/s which agrees very well.
From the difference of these values and considering only the ground frequency, that is
\begin{equation}
u' = \Delta\Omega_{max}\sigma \cos{(\omega t + \phi)},
\end{equation}
where $\phi$ is a phase angle. We obtain $u' = 6\cdot10^{-3}\cos{(\omega t + \phi)}$.
From Eq.~(\ref{eq:udecompose}) it appears that the particle ``sees'' a constant velocity in the negative y-direction and a small fluctuation on top of that.
For the calculation, with help of Eq.~(\ref{eq:rphiddot}) we therefore approximate $C_D$ and $C_L$ with the values based on $<u>$ with $\phi_e$=0. These are
\begin{subequations}
\begin{equation}
C_D = \frac{8}{3}\frac{gR}{(\omega r_e)^2}\frac{\rho_p-\rho}{\rho}
\end{equation}
\begin{equation}
C_L = \frac{1}{2}(1+C_A).
\end{equation}
\end{subequations}
With these values for $C_D$ and $C_L$ we calculated the particle trajectories for a number of frequencies in the suspended regime 0.07 Hz $<f_d<$ 0.11 Hz.
We used the value of $\chi$ to fit the calculated results to the measured ones.
As comparison of the results in Fig.~\ref{fig:NonRotatingParticles} with those in Fig.~\ref{fig:ODE} shows, in this way an excellent agreement is found.
The value of $\chi$ is 5$\cdot$10$^{-4}$.

Next we discuss the fact that the particle does not spin much around its axis.
In order to make an estimate for the expected spin, which can only be induced by the fluctuation part of the velocity, we first look at the relaxation time.
\citet{Bagchi2002} determined relaxation time for rotation of spheres in shear flow at Reynolds numbers up to 200 and found values about 4 times the quantity $d^2/\nu$, where $d$ is the particle diameter and $\nu$ is the kinematic viscosity of the liquid.
It is known that the relaxation time decreases with increasing Reynolds numbers.
We have Reynolds numbers of order 10$^3$ and therefore estimate the relaxation time $\tau$ as
\begin{equation}
\tau = \frac{d^2}{\nu} \sim 200 s
\end{equation}

This is much longer than the drum period,
\begin{equation}
\tau \gg \frac{2\pi}{\omega}.
\end{equation}

\begin{figure}
	\includegraphics[width=0.6\textwidth]{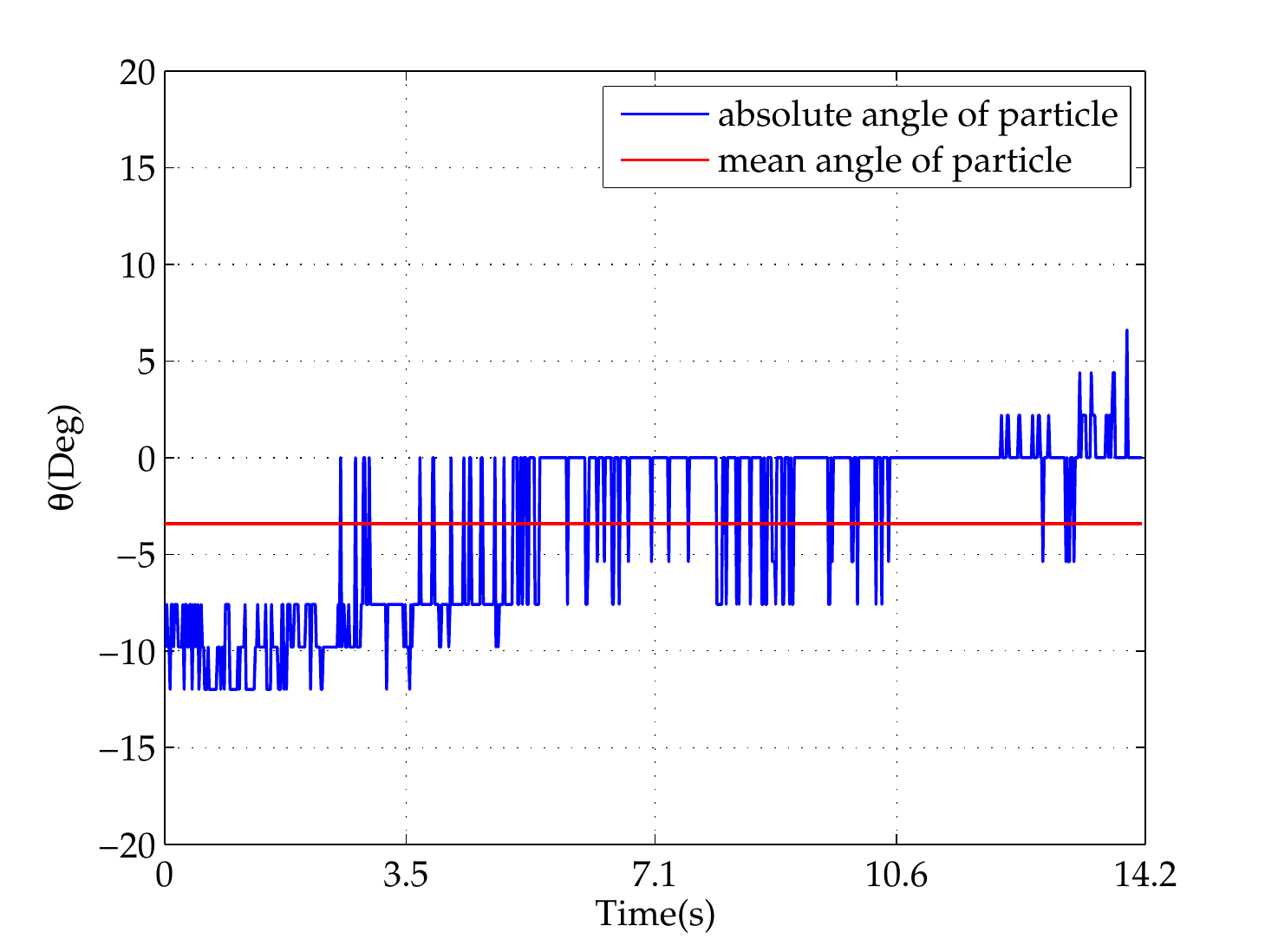}
	\caption{The absolute angular rotation of the sphere in a
solid body rotation flow with $f_d$ = 0.07 Hz for one period of the drum.}
	\label{fig:absrotation}
\end{figure}

A model equation for the induced rotation is
\begin{equation}
I\ddot{\theta}+F\dot{\theta} = M,
\end{equation}
where $I$ is the moment of inertia of the sphere: $I=2mR^2/5$ with mass $m=4\pi\rho_pR^3/3$.
Then, dividing by $I$, $F/I$ is 1/$\tau$, and the equation becomes
\begin{equation}
\ddot{\theta}+\frac{\dot{\theta}}{\tau} = \frac{M}{I}.
\end{equation}
We have not found expressions for the moment in an oscillating flow.
In general the moment exerted by the surrounding flow is expressed as
\begin{equation}
M = \frac{1}{2}C_M\rho(u')^2\pi R^3.
\label{eq:M}
\end{equation}
\citet{Zeng2005} report for a sphere in a shear flow and in the presence of a wall $C_M$ values of 10$^{-2}$-10$^{-3}$.
We take $C_M$= 0.01, insert this in Eq.~(\ref{eq:M}).
Using also the above expression for $I$ results in
\begin{equation}
\ddot{\theta}+\frac{\dot{\theta}}{\tau} \sim 10^{-2}e^{i\omega t}.
\end{equation}
Taking $\theta = Be^{i\omega t}$, we find for the real part of $B$
\begin{equation}
Re(B) = \frac{10^{-2}\tau^2}{1+(\omega\tau)^2}\sim\frac{10^{-2}}{\omega}.
\end{equation}
With our $\omega = 2\pi f_d = 0.69$ s$^{-1}$ this amounts to an oscillation of 1.2 degrees.
An oscillation of this order is observed indeed as Fig.~\ref{fig:absrotation} shows.

As described above, the orbits from Eq.~(\ref{eq:rphiddot}) with the wall repulsive force model can reproduce the experimental results. 
This suggests that even though the wall force model (Eq.~\ref{eq:F_W}) by \citet{Takemura2003} was made for a rising sphere near a vertical planar wall, it seems to be valid for the case of solid-body rotation in the present $Re$ regime.

\begin{figure}
	\centering
	\subfloat[Flow field at $f_d$ = 0.07Hz.]{\label{fig:Vectorplot007Hz}\includegraphics[width=0.5\textwidth]{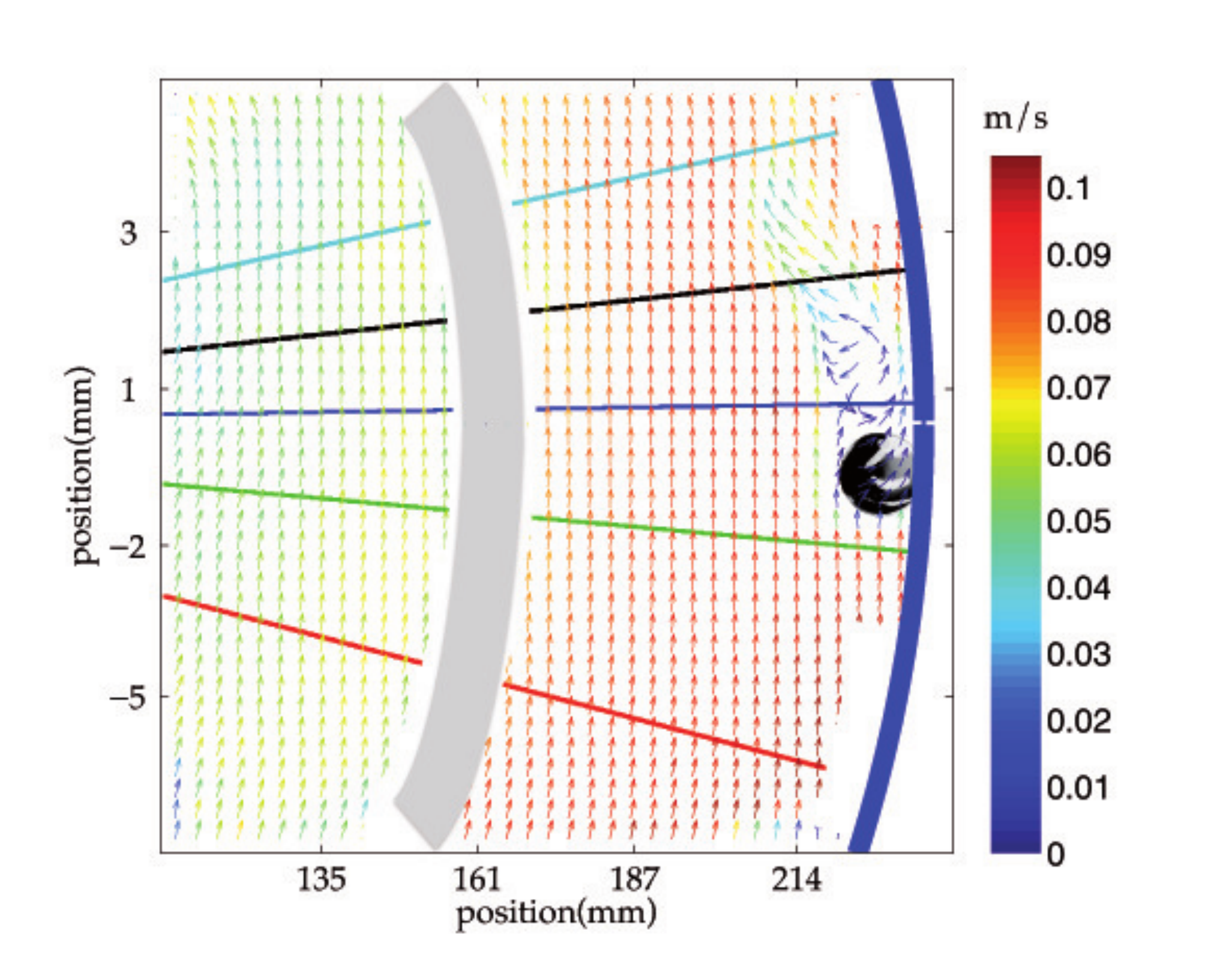}}
	\subfloat[Velocities vs. positions along the lines shown in the left figure.]{\label{fig:VThetaalongthetadiff}\includegraphics[width=0.5\textwidth]{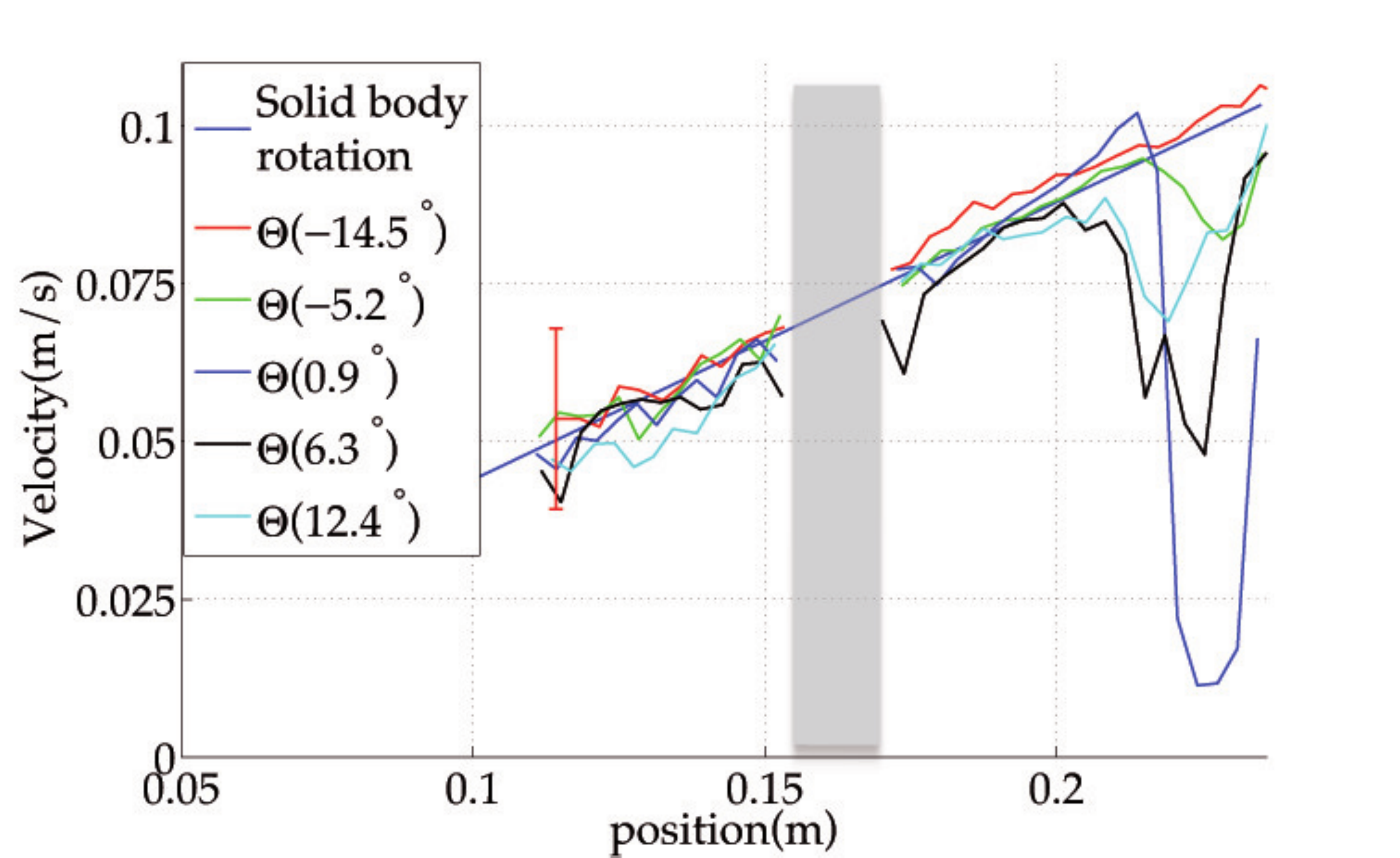}}                                
	\caption{The result of the flow field around a 7 mm particle in a drum at $f_d$=0.07 Hz using Particle Image Velocimetry.}	
	\label{fig:PIV}
\end{figure}

In addition, we performed PIV measurements.
A snapshot of the flow field is shown in Fig.~\ref{fig:PIV}(a).
The inward flow is undisturbed and the wake of a particle is deflected inwards the cylinder. 
This is more clearly observed in Fig.~\ref{fig:PIV}(b).
The upstream flow at $\theta=-14.5^{\circ}$ of the particle is undisturbed solid body rotation.
The wake effect is observed in downstream of the particle at $\theta=0.9^{\circ}$, $\theta=6.3^{\circ}$ and $\theta=12.4^{\circ}$.

\section{Conclusions}
\label{sec:concl}

We study translational and rotational motion of a slightly heavy particle in a rotating drum filled with water.
Remarkably in the regime of 0.07 Hz $\leq f_{drum} \leq$ 0.11 Hz the particle is found to be suspended in the drum with an orbital motion, which was not observed by \citet{Ashmore2005}. 
We investigate the force balance by including the effect of the drum wall ($F_W$)  modeled as $F_W \propto L^{-4}$ \cite{Takemura2003, Zeng2005}. 
The orbits of the particle from the force balance including $F_W$ can reproduce the experimental trajectories.
We also investigate the spin of a particle. 
The particle orientation changes less than 10 degrees over one cycle of the drum due to the small torque acting on the sphere. 

\section*{Acknowledgments}
We acknowledge support from the EU COST Action MP0806 on ``Particles in Turbulence''.
We thank Devaraj van der Meer and Detlef Lohse for fruitful discussions.

%
%

\bibliographystyle{apsrev4-1.bst}

\end{document}